\documentstyle[12pt]{article}

\begin{document}

\def\sfrac#1#2{{\textstyle{#1 \over #2}}}

\def\note#1{\thinspace$^{\#}$\footnote{\samepage #1}}

\def\agt{\hbox{${\lower.40ex\hbox{$>$}
\atop \raise.20ex\hbox{$\sim$}}$}}
\def\alt{\hbox{${\lower.40ex\hbox{$<$}
\atop \raise.20ex\hbox{$\sim$}}$}}

\hfill SFU HEP preprint 104-93 

\hfill July 1993

\vskip 0.5 in

\begin{center}
\large\bf $\Upsilon$ decay into charmonium \\
and the color octet mechanism
\end{center}

\vskip 0.5 in

\begin{center}
Howard D. Trottier \\ 
\it Department of Physics, \\[-6 pt]
Simon Fraser University, \\[-6 pt]
Burnaby, B.C., Canada V5A 1S6
\end{center}

\vskip 0.5 in

\centerline{\bf Abstract}
\vskip 12 pt

\noindent
A factorization theorem for $P$-wave quarkonium production,
recently derived by Bodwin, Braaten, Yuan and Lepage, is
applied to $\Upsilon \to \chi_{cJ} + X$, where $\chi_{cJ}$
labels the ${}^3 P_J$ charmonium states. The widths for
$\chi_{cJ}$ production through color-singlet $P$-wave 
and color-octet $S$-wave $c \bar c$ subprocesses are computed
each to leading order in $\alpha_s$. Experimental data on 
$\Upsilon \to J / \psi + X$ is used to obtain 
an upper bound on a nonperturbative parameter (related to the
probability for color-octet $S$-wave $c \bar c$ hadronization
into $P$-wave charmonium) that enters into the 
factorization theorem. 
The bound obtained here adds to the limited information so 
far available on the color-octet mechanism for 
$P$-wave quarkonium production.

\newpage

Factorization theorems play a basic role in perturbative QCD
calculations of many hadronic processes. A well known factorization
theorem for the decay and production of $S$-wave quarkonium
follows from a nonrelativistic description of heavy quark-antiquark
($Q \overline Q$) binding \cite{Kwong}. Nonperturbative effects are 
factored into $R_S(0)$, the nonrelativistic wave function at the origin, 
leaving a hard $Q \overline Q$ subprocess matrix 
element that can be calculated in perturbation theory. This 
factorization is valid to all orders in the strong coupling 
$\alpha_s$, and to leading order in $v^2$, where $v$ is 
the typical center-of-mass velocity of the heavy quarks.

Remarkably, the correct factorization theorems for 
the decay \cite{BBL} and production \cite{BBYL} of $P$-wave
quarkonium have only recently been derived. These new
theorems resolve a long standing problem regarding
infrared divergences which appear in some cases to leading order
in the rates for $P$-wave $Q \overline Q$ states \cite{Barbieri}.
In previous phenomenological calculations, the divergence was
replaced by a logarithm of a soft binding scale, such as the
binding energy or confinement radius \cite{Barbieri,Kwong}.
However, a rigorous calculation requires that one consider
additional components of the Fock space for $P$-wave quarkonium,
such as $\vert Q \overline Q g \rangle$, where
the $Q \overline Q$ pair is in a color-octet $S$-wave state, 
and $g$ is a soft gluon \cite{BBL,BBYL}. 

A renewed study of the decay and production of $P$-wave quarkonium 
is therefore of considerable interest, since one may gain new
information on a nonperturbative sector of QCD that has largely
been neglected in the quark model description of heavy quarkonium.
This is also of practical consequence; for example, $J / \psi$ 
production provides a clean experimental signature for many
important processes, and $P$-wave charmonium states have 
appreciable branching fractions to $J / \psi$.

In this paper the factorization theorem for $P$-wave quarkonium
production is applied to $\Upsilon \to \chi_{cJ} + X$, where 
$\chi_{cJ}$ labels the ${}^3 P_J$ charmonium states. The widths 
for $\chi_{cJ}$ production through color-singlet $P$-wave 
and color-octet $S$-wave $c \bar c$ subprocesses are computed
each to leading order in $\alpha_s$. Experimental data on 
$\Upsilon \to J / \psi + X$ is used to obtain 
an upper bound on a nonperturbative parameter (related to the
probability for color-octet $S$-wave $c \bar c$ hadronization
into $P$-wave charmonium) that enters into the 
factorization theorem. 
The bound obtained here adds to the limited information so 
far available on the color-octet mechanism for $P$-wave 
quarkonium production.
The color-octet component in $P$-wave 
decay was estimated in Ref. \cite{BBL} from measured
decay rates of the $\chi_{c1}$ and $\chi_{c2}$. A rough estimate
of the color-octet component in $P$-wave charmonium production 
was obtained in Ref. \cite{BBYL} from data on $B$ meson
decays; however, an accurate determination in that case
requires a calculation of next-to-leading order QCD corrections 
to the color-singlet component of $B \to \chi_{cJ} + X$,
which is so far unavailable \cite{BBYL}.

The factorization theorem for $P$-wave quarkonium production
has two terms, and in the case of $\Upsilon$ decay takes the form:
\begin{eqnarray}
   \Gamma(\Upsilon \to \chi_{cJ} + X)
   & = & H_1 \hat 
   \Gamma_1(\Upsilon \to c \bar c ( {}^3 P_J ) + X ; \mu)
\nonumber \\ 
   & & \mbox{} + (2 J + 1) H_8'(\mu) 
   \hat \Gamma_8(\Upsilon \to c \bar c ( {}^3 S_1 ) + X) .
\label{Fact}
\end{eqnarray}
$\hat \Gamma_1$ and $\hat \Gamma_8$ are hard subprocess rates
for the production of a $c \bar c$ pair in color-singlet $P$-wave 
and color-octet $S$-wave states respectively. The quarks 
are taken to have vanishing relative momentum. 
The nonperturbative parameters $H_1$ and $H_8'$ are proportional 
to the probabilities for these $c \bar c$ configurations to hadronize 
into a color-singlet $P$-wave bound state. $H_1$, 
$H_8'$ and $\hat \Gamma_8$ are independent of 
the total angular momentum $J$.

This factorization theorem is valid to all orders in $\alpha_s$
and to leading order in $v^2$. The hard subprocess rates are free 
of infrared divergences. $\hat \Gamma_1$ and $H'_8$ depend on an 
arbitrary factorization scale $\mu$ in such a way that the physical
decay rate is independent of $\mu$. In order to avoid large
logarithms of $m_\Upsilon / \mu$ in $\hat \Gamma_1$,
$\mu$ of $O(m_\Upsilon)$ should be used.
The factorization theorem for $P$-wave charmonium decay 
contains a nonperturbative parameter $H_8$ analogous to $H_8'$. 
However a production process must be used to determine $H_8'$ 
phenomenologically \cite{BBYL}. 

In the usual nonrelativistic quark model, $H_1$ can be expressed 
in terms of the $P$-wave color-singlet $c \overline c$  
wave function:
\begin{equation}
   H_1 \approx {9 \over 2\pi} 
               {\vert R_P'(0) \vert^2 \over m_c^4} 
   \approx 15~{\rm MeV} ,
\label{H1}
\end{equation}
where the numerical estimate was obtained in Ref. \cite{BBL}
from measured decay rates of the $\chi_{c1}$ and $\chi_{c2}$.

$H_8'$ cannot be rigorously expressed perturbatively in terms 
of $R_P$, since it accounts for radiation of a soft gluon by 
a color-octet $c \bar c$ pair. 
The scale dependence of $H_8'(\mu)$ is determined by the 
following renormalization-group equation 
(to leading order in $\alpha_s(\mu)$) \cite{BBL,BBL2}:
\begin{equation}
   \mu {d \over d\mu} H_8'(\mu) \approx
   {16 \over 27\pi} \alpha_s(\mu) H_1 ,
\label{RGE}
\end{equation}
which is readily integrated. For example \cite{BBYL}:
\begin{equation}
   H_8'(m_b) = H_8'(\mu_0) + \left[ 
   {16 \over 27 \beta_3} 
   \ln\left( {\alpha_s(\mu_0) \over \alpha_s(m_c)} \right)
 + {16 \over 27 \beta_4}
   \ln\left( {\alpha_s(m_c) \over \alpha_s(m_b)} \right)
   \right] H_1 
\label{H8Mb}
\end{equation}
(for $\mu_0 < m_c$), where $\beta_n = (33 - 2n) / 6$.
If $H_8'(\mu_0)$ is neglected in the limit of large $m_b$
one obtains $H_8'(m_b) \approx 3$~MeV,
using $\alpha_s(\mu_0) \sim 1$ \cite{BBYL}. While one might
not expect the physical value of $m_b$ to be large enough 
to neglect $H_8'(\mu_0)$, an estimate for $H_8'(m_b)$ 
obtained in Ref. \cite{BBYL} from experimental data on 
$B$ meson decays is consistent with the above result. 

A calculation of $\hat \Gamma_1$ and $\hat \Gamma_8$
in Eq. (\ref{Fact}) each to leading order in $\alpha_s$ 
can be obtained from a calculation of the infrared divergent width 
$\Gamma_{\rm div}$ for $\Upsilon \to c \bar c ({}^3 P_J) + ggg$,
where the $c \bar c$ pair is in a color-singlet $P$-wave state:
\begin{eqnarray}
\lefteqn{
   \Gamma_{\rm div}
   (\Upsilon \to c \bar c ({}^3 P_J) + ggg; \mu_0) 
   \equiv
        }
\nonumber \\ 
   & & {20 \alpha_s^5 \over 3^7 \pi^3} 
   {G^\Upsilon_1 \over m_\chi}
   \left[ {\cal F}_{1J}(\mu) + 
   (2J + 1) {16 \over 27\pi} 
   \ln\left( {\mu \over \mu_0} \right) {\cal F}_8 \right] H_1 . 
\label{Gdiv}
\end{eqnarray}
${\cal F}_{1J}$ and ${\cal F}_8$ are dimensionless infrared-finite 
form factors. $\mu_0$ is an infrared cutoff on the energy of 
soft gluons, and $\mu$ is an arbitrary factorization scale
[the $\mu$ dependence of ${\cal F}_{1J}$ exactly cancels that
of the explicit logarithm in Eq. (\ref{Gdiv})].

The constants in Eq. (\ref{Gdiv}) include a color-factor of $5 / 216$ 
and phase space factors, including $1/3$ for $\Upsilon$ spin-averaging,
and $1/3!$ for the phase space of the three indistinguishable gluons
[cf. Eq. (\ref{Phase}) below].
$G^\Upsilon_1$ is related to the usual $S$-wave $b \bar b$ 
nonrelativistic wave function:
\begin{equation}
   G^\Upsilon_1 \approx {3 \over 2\pi}
   {\vert R^\Upsilon_S(0) \vert^2 \over m_b^2}
   \approx 108~{\rm MeV} ,
\label{G1}
\end{equation}
where the numerical value is obtained from the electronic decay
rate of the $\Upsilon$ \cite{PDG}.

The hard subprocess rates of Eq. (\ref{Fact}) are identified
from $\Gamma_{\rm div}$ by using the perturbative expression
for the infrared divergence in $H_8'$, obtained from Eq. (\ref{RGE}) 
by neglecting the running of the coupling \cite{BBYL}
\begin{equation}
   H_8'(\mu) \sim {16 \over 27\pi} \alpha_s
   \ln\left( {\mu \over \mu_0} \right) H_1 .
\label{H8H1}
\end{equation}
Thus:
\begin{equation}
   \hat \Gamma_1(\Upsilon \to c \bar c ({}^3 P_J) + ggg; \mu)
   = {20 \alpha_s^5 \over 3^7 \pi^3} 
     {G^\Upsilon_1 \over m_\chi} 
     {\cal F}_{1J}(\mu) ,
\label{hat1}
\end{equation}
and
\begin{equation}
   \hat \Gamma_8(\Upsilon \to c \bar c ({}^3 S_1) + gg)
   = {20 \alpha_s^4 \over 3^7 \pi^3} 
     {G^\Upsilon_1 \over m_\chi} 
     {\cal F}_8 .
\label{hat8}
\end{equation}
Note that $\hat\Gamma_1$ is suppressed by $O(\alpha_s)$
compared to $\hat\Gamma_8$. However, the nonperturbative 
parameters $H_1$ and $H_8'$ which accompany these subprocess
rates in Eq. (\ref{Fact}) are independent, hence $\alpha_s H_1$
need not be small compared to $H_8'$ \cite{BBL,BBYL}.
We therefore proceed to calculate $\hat\Gamma_1$
and $\hat\Gamma_8$ each to leading order; all further 
corrections to $P$-wave production are then guaranteed
to be suppressed by at least one power of $\alpha_s$
compared to what is included here. 

In order to extract ${\cal F}_{1J}$ and ${\cal F}_8$ individually,
it is necessary to explicitly identify the infrared logarithm in 
the calculation of $\Gamma_{\rm div}$. This can be done 
analytically, as described in the following.

There are 36 $O(\alpha_s^5)$ diagrams contributing to 
$\Gamma_{\rm div}$. One of these is shown in 
Fig. \ref{FigFeynman}. All infrared divergences
are associated with a gluon that is radiated from a charm 
quark line; in Fig. \ref{FigFeynman} this gluon 
carries four momentum $k_1$ and polarization $\epsilon_1$.
Define the invariant amplitude ${\cal M}_J(2,3;1)$
corresponding to the sum of all Feynman diagrams
where gluon ``1'' is radiated from the charm quark line.
The amplitude is readily computed using expressions 
for $S$- and $P$-wave $Q \overline Q$ currents given in 
Ref. \cite{Kuhn}%
\note{Overall factors in the quark currents
including couplings, color amplitudes, and wave functions
have been accounted for in Eq. (\ref{Gdiv}).}
\begin{equation}
   {\cal M}_J(2,3;1) \equiv -
   { m_\Upsilon m_\chi  B_\mu(2,3)  C_J^\mu(1)
   \over 
   [ (k_2 + k_4) \cdot k_3 ] [ (k_3 + k_4) \cdot k_2 ]
   [ (k_2 + k_3) \cdot k_4 ] \, k_4^2 \, (k \cdot k_1)^2 } ,
\label{M231}
\end{equation}
where 
\begin{eqnarray}
   \epsilon_4^\mu B_\mu(2,3) & = &
   \bigl\{ 
   \epsilon_4 \cdot \epsilon_2 [ 
     - k_4 \cdot k_3 \epsilon_3 \cdot k_2 \epsilon_0 \cdot k_4
     - k_2 \cdot k_3 \epsilon_3 \cdot k_4 \epsilon_0 \cdot k_2
     - k_4 \cdot k_3 k_2 \cdot k_3 \epsilon_0 \cdot \epsilon_3 ]
\nonumber \\ 
   & & \mbox{} + 
   \epsilon_0 \cdot \epsilon_3 [
       k_4 \cdot k_3 \epsilon_4 \cdot k_2 \epsilon_2 \cdot k_3
     + k_2 \cdot k_3 \epsilon_2 \cdot k_4 \epsilon_4 \cdot k_3
     - k_4 \cdot k_2 \epsilon_4 \cdot k_3 \epsilon_2 \cdot k_3 ]
   \bigr\}
\nonumber \\ 
   & & \mbox{} + \left\{ 2 \leftrightarrow 3 \right\}
               + \left\{ 3 \leftrightarrow 4 \right\} 
\label{B23}
\end{eqnarray}
($\epsilon_0$ is the polarization of the $\Upsilon$), and
\begin{eqnarray}
   & & \epsilon_{4\mu} C_{J=0}^\mu(1) =
   \sqrt{\sfrac16} 
   \left[ \epsilon_1 \cdot \epsilon_4 k_1 \cdot k_4
        - \epsilon_1 \cdot k_4 \epsilon_4 \cdot k_1 \right]
   \left( m_\chi^2 + k \cdot k_4 - k_4^2 \right) ,
\nonumber \\ 
   & & \epsilon_{4\mu} C_{J=1}^\mu(1) =
   \sfrac12 m_\chi k_4^2 \,
   \varepsilon_{\alpha\beta\gamma\delta} \,
   e^\alpha \epsilon_4^\beta \epsilon_1^\gamma k_1^\delta ,
\label{C1} \\ 
   & & \epsilon_{4\mu} C_{J=2}^\mu(1) =
   \sqrt{\sfrac12} m_\chi^2 \left(
   k_1 \cdot k_4 \epsilon_1^\alpha \epsilon_4^\beta  +  
   k_4^\alpha k_1^\beta \epsilon_1 \cdot \epsilon_4  -
   k_1^\alpha \epsilon_4^\beta \epsilon_1 \cdot k_4  -
   k_4^\alpha \epsilon_1^\beta \epsilon_4 \cdot k_1 \right) 
   e^{\alpha\beta} .
\nonumber
\end{eqnarray}
$e^\alpha$ is a spin-1 polarization vector 
and $e^{\alpha\beta}$ is a spin-2 polarization tensor.

For convenience the virtual gluon is labeled in 
Eqs. (\ref{M231})--(\ref{C1}) by polarization $\epsilon_4$ 
and momentum $k_4$ ($k_4 = P - k_2 - k_3 = k + k_1$).
Terms which vanish due to the on-shell conditions
$\epsilon_i \cdot k_i = 0$ ($i=1,2,3$) and 
$\epsilon_0 \cdot P = 0$ have been dropped.
The charm quark current $C_J^\mu(1)$ is
symmetric under interchange of labels 1 and 4 
(up to terms which vanish due to the on-shell conditions).
The bottom quark current $B_\mu(2,3)$ is explicitly
symmetric under interchange of labels $2$, $3$ and $4$.

The overall factors in Eq. (\ref{Gdiv}) are such that:
\begin{eqnarray}
\lefteqn{
   {\cal F}_{1J}(\mu) + 
   (2J + 1) {16 \over 27\pi} 
   \ln\left( {\mu \over \mu_0} \right) {\cal F}_8
   \equiv 
        }
\nonumber \\ 
   & & 3 \int d[\Phi_4] \sum_{\rm spins} 
   \left[ {\cal M}_J^2(2,3;1) 
      + 2 {\cal M}_J(2,3;1) {\cal M}_J(1,3;2) \right] ,
\label{Phase}
\end{eqnarray}
where $\Phi_n$ denotes (infrared-cutoff) $n$-body phase space, 
normalized according to
\begin{equation}
   \Phi_n[P \to p_1,\ldots,p_n] \equiv
   \int \prod_{i=1}^n {d^3p_i \over 2E_i}
   \delta^4(P - \sum_i p_i) .
\label{Phi}
\end{equation}
The factor of 3 on the right hand side of Eq. (\ref{Phase}) 
accounts for symmetrization of ${\cal M}_J(2,3;1)$
under gluon label interchanges $1 \leftrightarrow 2$
and $1 \leftrightarrow 3$, taking account
of the symmetry in the three gluon phase space.

The infrared divergence comes entirely from the first 
term in square brackets in Eq. (\ref{Phase}), 
and is due to the $P$-wave
charm quark propagator $1/(k.k_1)^2$ in Eq. (\ref{M231}).
It is therefore advantageous to organize the four-body
phase space integral in Eq. (\ref{Phase}) by taking 
the invariant mass of the $\chi_{cJ}$ and gluon ``1'' as one
integration variable \cite{Byckling}
\begin{eqnarray}
\lefteqn{
   \int d[\Phi_4] = 
   \int_0^{(m_\Upsilon - m_\chi)^2} 
        d(k_{23}^2) \,
   \int_{m_\chi^2 + 2 \mu_0 m_\chi}^{(m_\Upsilon - m_{23})^2} 
        d(k_{1\chi}^2) \,
        }
\nonumber \\
   & & \mbox{} \times
       \Phi_2 [ P \to k_{23} , k_{1\chi} ] \,
       \Phi_2 [ k_{23} \to k_2 , k_3 ] \,
       \Phi_2 [ k_{1\chi} \to k_1 , k ] ,
\label{Phi4}
\end{eqnarray}
where $m_{23}^2 \equiv k_{23}^2$.
Note the infrared cutoff $\mu_0$ on the energy of gluon ``1''
in the rest frame of the $\chi_{cJ}$.

The infrared logarithm on the right-hand side of Eq. (\ref{Phase})
can now be identified analytically by observing that
$B_\mu(2,3) C^\mu_J(1)$ in Eq. (\ref{M231}) is given by a 
sum of terms each containing exactly one factor of $k_1$, if
$k_4 = P - k_2 - k_3$ is used to eliminate the virtual gluon 
momentum. With this convention, one has
\begin{equation}
   \sum_{\rm spins} {\cal M}_J^2(2,3;1) =
   {\gamma_J(k_1; P, k, k_2, k_3) \over (k \cdot k_1)^2} ,
\label{gk1}
\end{equation}
where $k_1$ appears explicitly in the function 
$\gamma_J(k_1; P, k, k_2, k_3)$ only in the combination
$k_1 / k \cdot k_1$. 

${\cal F}_8$ is then given in terms of a manifestly 
infrared-finite three-body phase space integral, taking account 
of the fact that 
$\Phi_2( k_{1\chi} \to k_1 , k ) 
= \sfrac14 k \cdot k_1 / k_{1\chi}^2 \int d\Omega_{1\chi}^*$,
where $\Omega_{1\chi}^*$ is the center-of-mass solid
angle of the two body system:
\begin{eqnarray}
   (2J + 1) {\cal F}_8 & = & 
   {27 \pi \over 32 m_\chi^2}
   \int_0^{(m_\Upsilon - m_\chi)^2} 
       d(k_{23}^2) \,
      \Phi_2 [ P \to k_{23} , k ]
\nonumber \\ 
   & & \mbox{} \times 
      \Phi_2 [ k_{23} \to k_2 , k_3 ]
   \int d\Omega_{1\chi}^* \, 
      \gamma_J(\widetilde k_1; P, k, k_2, k_3) ,
\label{F8int}
\end{eqnarray}
where
\begin{equation}
   \widetilde k_1 \equiv \lim_{k \cdot k_1 \to 0} 
   {k_1 \over k \cdot k_1} .
\label{k1soft}
\end{equation}
The finite four-vector $\widetilde k_1$ is readily expressed 
directly in terms of $k_{23}^2$ and $\Omega_{1\chi}^*$.
An expression for ${\cal F}_{1J}$ can be obtained 
from Eqs. (\ref{Phase}) and (\ref{F8int}) by analogy with
the identity
$\int dx f(x) / x = f(0) \ln x + \int dx [f(x) - f(0)] / x$.

The contraction of currents and sum over polarizations 
in Eqs. (\ref{M231}) and (\ref{Phase}) were performed 
symbolically using {\small REDUCE} \cite{REDUCE} 
(leading to lengthy expressions, particularly for $J=2$). 
The $\chi_{cJ}$ spin sums were done using 
(see e.g. Ref. \cite{Kuhn}):
\begin{eqnarray}
   & & \sum_e e_\mu e_\nu = - g_{\mu\nu} 
    + {k_\mu k_\nu \over m_\chi^2} \equiv {\cal P}_{\mu\nu} ,
\nonumber \\ 
   & & \sum_e e_{\mu\nu} e_{\alpha\beta}
    = \sfrac12 \left[ 
      {\cal P}_{\mu\alpha} {\cal P}_{\nu\beta}
    + {\cal P}_{\mu\beta}  {\cal P}_{\nu\alpha} \right]
    - \sfrac13
      {\cal P}_{\mu\nu} {\cal P}_{\alpha\beta} .
\label{chipol}
\end{eqnarray}
The phase space integrals were evaluated numerically using 
{\small VEGAS} \cite{VEGAS}; modest integration grids are 
found to give very good convergence.

The fact that ${\cal F}_8$ should be independent of $J$
provides a stringent check of these calculations, given that
the three currents $C_J^\mu$ have very different structures
[cf. Eq. (\ref{C1})]. This was verified explicitly in 
numerical calculations of Eq. (\ref{F8int}), to better
than a few tenths of a percent for all $m_\chi / m_\Upsilon$
on a modest integration grid.

Figure \ref{FigF8} shows the numerical results for ${\cal F}_8$
over a range of hypothetical meson masses.
In Fig. \ref{FigF1} results for ${\cal F}_{1J}(\mu)$ are shown 
using a factorization scale $\mu = m_\Upsilon$.

The available experimental data on charmonium production
in $\Upsilon$ decay is for the $J / \psi$:
\begin{equation}
   B_{\rm exp} (\Upsilon \to J / \psi + X)
   \ \left\{
   \begin{array}{ll}
   = (1.1 \pm 0.4) \times 10^{-3} 
       & \mbox{{\small CLEO} \cite{CLEO},}    \\
   <  1.7 \times 10^{-3}
       & \mbox{Crystal Ball \cite{Crystal},}  \\ 
   <  0.68 \times 10^{-3}
       & \mbox{{\small ARGUS} \cite{ARGUS}.}
   \end{array}
   \right.
\label{Bexp}
\end{equation}
An upper bound on $H_8'$ can be extracted from this data
by computing the ``indirect'' production 
of $J / \psi$ due to the $\chi_{cJ}$ states.
Assuming that radiative cascades from $\chi_{c1}$ 
and $\chi_{c2}$ dominate, with branching fractions
$B_{\rm exp}(\chi_{c1} \to \gamma J / \psi) \approx 27\%$
and
$B_{\rm exp}(\chi_{c2} \to \gamma J / \psi) \approx 13\%$
\cite{PDG}, the results presented here give:
\begin{equation}
   H_8'(m_\Upsilon) \approx \left\{ 
   { \sum_J B(\Upsilon \to \chi_{cJ} + X' \to J / \psi + X) 
     \over 2.9 \times 10^{-5} }
   + 1.4 \right\} \mbox{MeV} .
\label{H8Bpsi}
\end{equation}
The first number in brackets above comes from the
color-octet subprocess rate $\hat \Gamma_8$, and the
second number from the color-singlet rate $\hat \Gamma_1$.
The experimental value for the total width
$\Gamma_{\rm tot}(\Upsilon) \approx 52$~keV \cite{PDG}
was used, along with 
$\alpha_s(m_\Upsilon) \approx 0.179$ \cite{Kwong},
and the values of $H_1$ and $G^\Upsilon_1$
given in Eqs. (\ref{H1}) and (\ref{G1}).

Equation (\ref{H8Bpsi}) yields the bound 
$H_8'(m_\Upsilon) \alt 25$~MeV
using the {\small ARGUS} upper limit, which is consistent 
with the other measurements. This bound is considerably larger
than an estimate $H_8'(m_b) \approx 3$~MeV
based on $B$ meson decays \cite{BBYL},%
\note{From Eq. (\ref{RGE}), $H_8'(\mu)$ increases by only 
$\approx 0.3$~MeV in the evolution
from $\mu=m_b$ to $\mu=m_\Upsilon$.}
although a calculation of next-to-leading order 
QCD corrections to the color-singlet component of
$B \to \chi_{cJ} + X$ is required before an accurate
determination of $H_8'$ can be made in that case \cite{BBYL}.

This raises the possibility of significant direct production 
of $J / \psi$ in the decay of the $\Upsilon$, 
unless the branching fraction turns out to be
considerably smaller than the {\small ARGUS} bound. Mechanisms 
for direct $\Upsilon \to J / \psi + X$ in perturbative QCD were first 
discussed in Refs. \cite{Fritzsch} and \cite{Bigi}. The direct
production rate is suppressed by $O(\alpha_s^2)$ compared to the
$P$-wave color-octet production mechanism considered here.
However, the nonperturbative matrix elements which enter into 
$P$-wave production are of $O(v^2)$ relative to the corresponding 
parameter for $S$-wave production, where $v$ is a typical
relative velocity of the quarks. Moreover, there are many 
channels which contribute to direct production.

The full $O(\alpha_s^6)$ perturbative QCD amplitude for 
direct $\Upsilon \to J / \psi + X$ was recently evaluated
in Ref. \cite{Irwin}, corresponding to one-loop diagrams
for $\Upsilon \to J / \psi + gg$, and tree diagrams
for $\Upsilon \to J / \psi + gggg$. 
The $O(\alpha_s^2 \alpha^2)$ electromagnetic amplitude
for the two gluon decay mode was also evaluated.
Unfortunately, only a crude estimate of the required phase space 
integrations was made in Ref. \cite{Irwin} (there is a costly 
convolution with a numerical calculation of the loop 
integrals for $\Upsilon \to J / \psi + gg$).
Nevertheless, the calculation of Ref. \cite{Irwin}
suggests a branching fraction for direct production
of a $\mbox{few} \times 10^{-4}$.
This would lead to a considerable reduction in the bound
on $H_8'$ extracted from Eqs. (\ref{Bexp}) and (\ref{H8Bpsi}).

To summarize, a complete calculation was made of the leading order
rates for $\Upsilon \to \chi_{cJ} + X$, through both color-singlet 
$P$-wave and color-octet $S$-wave $c \bar c$ subprocesses.
Experimental data on $J / \psi$ production 
was used to obtain an upper bound on the nonperturbative 
parameter $H_8'$, related to the probability for
color-octet $S$-wave $c \bar c$ hadronization into 
$P$-wave charmonium. 
This work adds to the limited information so far available on 
the color-octet mechanism for $P$-wave quarkonium 
decay and production \cite{BBL,BBYL}. These
investigations provide new information on a nonperturbative
sector of QCD that has largely been neglected in previous
studies of heavy quarkonium. A quantitative estimate of $H_8'$
is phenomenologically important since this parameter is required 
as input for the calculation of a variety of processes.
Improved experimental data, and a
definitive calculation of the direct $J / \psi$ production rate 
along the lines of Ref. \cite{Irwin}, would allow for an
accurate determination of $H_8'$ from the results presented here. 

\vskip 24 pt

I am indebted to Eric Braaten for suggesting this problem,
and for many enlightening conversations. I also thank 
Mike Doncheski, John Ng, and Blake Irwin for helpful 
discussions. This work was supported in part by the 
Natural Sciences and Engineering Research Council of Canada.


\begin{figure}[p]
\caption{One of the 36 $O(\alpha_s^5$) diagrams contributing
to $\Upsilon \to c \bar c ({}^3 P_J) + ggg$.\label{FigFeynman}}
\end{figure}

\begin{figure}[p]
\caption{Color-octet form factor ${\cal F}_8$ as a function
of $m_\chi / m_\Upsilon$.\label{FigF8}}
\end{figure}

\begin{figure}[p]
\caption{Color-singlet form factors ${\cal F}_{1J}$ as functions
of $m_\chi / m_\Upsilon$: $J=0$ (short-dashed line),
$J=1$ (long-dashed line), $J=2$ (solid line). The
form-factors were evaluated using a factorization scale
$\mu = m_\Upsilon$.\label{FigF1}}
\end{figure}

\end{document}